\documentclass[12pt,preprint]{aastex}

\shorttitle{Dynamical models of NGC$\,6397$}
\shortauthors{Hurley et al.}

\slugcomment{Submitted to the Astronomical Journal}

\begin{document}

\title{Deep ACS Imaging in the Globular Cluster NGC$\,6397$: Dynamical Models}

\author{Jarrod R. Hurley \\
\affil{Centre for Astrophysics and Supercomputing, 
        Swinburne University of Technology, \\
        P.O. Box 218, VIC 3122, Australia \\
{\tt jhurley@swin.edu.au}}}

\author{Michael M. Shara \\
\affil{Department of Astrophysics, 
        American Museum of Natural History, \\
       Central Park West at 79th Street, 
       New York, NY 10024 \\
{\tt mshara@amnh.org}}}

\author{Harvey B. Richer \\
\affil{Department of Physics and Astronomy, University of British Columbia, \\
        Vancouver, BC, Canada \\
{\tt richer@astro.ubc.ca}}}

\author{Ivan R. King \\
\affil{Department of Astronomy, University of Washington, Seattle, Washington \\
{\tt king@astro.washington.edu}}}

 \author{D. Saul Davis \\
\affil{Department of Physics and Astronomy, University of British Columbia, \\
        Vancouver, BC, Canada \\
{\tt sdavis@astro.ubc.ca}}}

\author{Jason S. Kalirai \\
\affil{Department of Astronomy, University of California at Santa Cruz, 
        Santa Cruz, California \\
{\tt jkalirai@ucolick.org}}}

\author{Brad M. S. Hansen \\
\affil{Department of Physics and Astronomy, University of California at Los Angeles, \\ 
        Los Angeles, California \\
{\tt hansen@astro.ucla.edu}}}

\author{Aaron Dotter \\
\affil{Department of Physics and Astronomy, Dartmouth College, 
        Hanover, New Hampshire \\
{\tt aaron.l.dotter@dartmouth.edu}}}

\author{Jay Anderson \\
\affil{Space Telescope Science Institute, Baltimore, Maryland \\
{\tt jayander@stsci.edu}}}

\author{Gregory G. Fahlman \\
\affil{HIA/NRC, Victoria, BC, Canada \\
{\tt greg-fahlman@nrc-cnrc.gc.ca}}}

\author{R. Michael Rich \\
\affil{Department of Physics and Astronomy,  \\
        Los Angeles, California\\
{\tt rmr@astro.ucla.edu}}}

\begin{abstract}
We present $N$-body models to complement deep imaging of the metal-poor core-collapsed 
cluster NGC$\,6397$ obtained with the Hubble Space Telescope. 
All simulations include stellar and binary evolution in-step with the stellar dynamics 
and account for the tidal field of the Galaxy.  
We focus on the results of a simulation that began with $100\,000$ objects (stars and binaries), 
5\% primordial binaries and Population II metallicity.  
After $16\,$Gyr of evolution the model cluster has about 20\% of the stars remaining and has 
reached core-collapse. 
We compare the color-magnitude diagrams of the model at this age for the central region 
and an outer region corresponding to the observed field of NGC$\,6397$ 
(about $2$--$3$ half-light radii from the cluster centre). 
This demonstrates that the white dwarf population in the outer region has suffered little 
modification from dynamical processes -- contamination of the luminosity function by binaries 
and white dwarfs with non-standard evolution histories is minimal and should not 
significantly affect measurement of the cluster age. 
We also show that the binary fraction of main-sequence stars observed in the NGC$\,6397$ 
field can be taken as representative of the primordial binary fraction of the cluster. 
For the mass function of the main-sequence stars we find that although this has been 
altered significantly by dynamics over the cluster lifetime, especially in the central and outer 
regions, that the position of the observed field is close to optimal for recovering the 
initial mass function of the cluster stars (below the current turn-off mass). 
More generally we look at how the mass function changes with radius in a dynamically evolved 
stellar cluster and suggest where the best radial position to observe the initial mass function 
is for clusters of any age. 
We discuss computational constraints that restrict the $N$-body method to non-direct models 
of globular clusters currently, how this affects the interpretation of our results regarding 
NGC$\,6397$, and future plans for models with increased realism. 
\end{abstract}


\section{Introduction}
\label{s:intro}

As part of a program to measure the white dwarf (WD) cooling ages of globular clusters 
a dataset of the nearby cluster NGC$\,6397$ has been obtained that is 
unprecedented in its depth (Richer et al. 2006). 
This involved imaging a single field with the Advanced Camera for Surveys (ACS) 
on the Hubble Space Telescope (HST) for 126 orbits during Cycle 13. 
The field in question is located 5 arcminutes SE of the cluster core and is 
complemented by simultaneous exposures of the cluster core obtained with 
the Wide Field Planetary Camera (WFPC2). 
Results obtained to date from this rich dataset include: 
observational confirmation of the theoretical prediction for the termination point of 
the stellar main-sequence at the low-mass end (Richer et al. 2006); 
presentation of the colour-magnitude diagram and analysis of the luminosity function 
of the cluster stars (Richer et al. 2007); 
determination of the WD cooling age (Hansen et al. 2007); 
a study of the radial distribution of the cluster WDs (Davis et al. 2008a); 
calculation of the space motion of the cluster (Kalirai et al. 2007); and, 
measurement of the binary fraction of NGC$\,6397$ (Davis et al. 2008b). 
In the majority of these publications the results of realistic $N$-body models of star cluster evolution 
have been utilised in varying degrees to ascertain the extent to which dynamical 
evolution has impacted the observed data. 
The function of this paper is to provide the necessary background to the $N$-body models 
and to look at the model results in more detail. 

NGC$\,6397$ is located in the inner halo of the Galaxy and has an age in the range of $11-13\,$Gyr 
(Pasquini et al. 2004; Hansen et al. 2007). 
It has previously been shown to have a ``collapsed'' core (Djorgovski \& King 1986), 
meaning that it has evolved past the potentially catastrophic phase of core-collapse 
and is dynamically old (Spitzer 1987). 
It has also been shown to exhibit mass segregation (King et al. 1998, for example). 
Mass-segregation is a direct consequence of the tendency towards equipartition 
of kinetic energies for stars in a bound stellar system. 
This means that heavier stars move more slowly on average and sink towards the 
center of the system, while low-mass stars move outwards in radius. 
The observed mass function (MF) of a star cluster will therefore be 
affected by mass segregation and the extent to which it is affected can be expected to 
depend on the prior dynamical evolution of the cluster 
-- as previously observed by Piotto \& Zoccali (1999). 

Methods used to quantify the effects of mass-segregation include 
equilibrium models, such as King (1966), and simulations that follow 
the dynamical evolution of a star cluster. 
The latter typically employ either the direct $N$-body method 
(Baumgardt \& Makino 2003, for example) or the statistical 
Monte Carlo (e.g. Fregeau et al. 2002) and Fokker-Planck (e.g. Drukier 1995) methods. 
While much can be learnt by studying the relatively ``clean'' problem of mass-segregation 
in a self-gravitating system with only two stellar mass groups 
(Khalisi, Amaro-Seoane \& Spurzem 2007) the effect on the MF of a cluster as it evolves 
is more directly understood by employing a distribution of masses. 
Giersz \& Heggie (1996) used $N$-body models starting with $500$ stars and masses 
distributed according to an initial mass function (IMF) to demonstrate that a state 
of true energy equipartition is never achieved in practice. 
The evolution of the MF was explored in detail by Vesperini \& Heggie (1997) through 
$N$-body models with $N \sim 4\,000$--$16\,000$ stars and including effects such as 
stellar evolution, the external tidal field of the Galaxy and disk-shocking. 
More recently this has been extended to models starting with as many as $131\,072$ 
stars by Baumgardt \& Makino (2003). 
In this work we study the MF of a dynamically old cluster using $N$-body models 
that start with $N = 100\,000$ and include primordial binaries 
(shown to be important by de la Fuente Marcos 1996). 
We extend the study by also looking at the expected color-magnitude diagram 
of such a cluster as well as the radial distributions of the binaries and WDs.  

The setup of our models and the simulation method are detailed in Section~2. 
This includes an initial discussion of issues 
related to timescales and scaling between the model and real data. 
An overview of the evolution history of our main model is also given in Section~2. 
Results are presented in Section 3 with a focus on NGC$\,6397$ and this is followed 
by a discussion in Section 4 of the broader implications and future modeling aims. 
We briefly summarize our findings in Section 5.
 
\section{The $N$-body Models}
\label{s:method}

\subsection{Overview of the $N$-body Code and Method} 

To investigate the long-term evolution of star clusters we use the 
Aarseth {\tt NBODY4} code \citep{aar99}. 
This is a brute force $N$-body code that directly integrates the 
$N$ individual equations of motion for the $N$ bodies (stars or binaries) 
that make up a cluster. 
Unlike galaxy-scale $N$-body simulations there is no softening involved 
in the force equation -- collisional dynamics is an essential aspect of 
star cluster evolution. 
Cumulative weak gravitational interactions drive dynamical relaxation 
which sets the timescale for processes such as mass-segregation 
and core-collapse. 
Stronger interactions can lead to binary formation and the ejection of 
stars from the cluster. 
Also important is the feedback between the stellar and binary evolution 
of the cluster components and the evolution of the cluster itself. 
For example, mass-loss from stars affects the gravitational potential 
of the cluster and must be accounted for in self-consistent simulations. 
Furthermore, star clusters do not evolve in isolation 
so the tidal field of the host galaxy must also be considered. 
The strength of the tidal field restricts the physical size of the cluster 
and also the dynamical evolution timescale. 

Traditionally the $N$-body method has been the preferred approach for producing 
realistic models of star cluster evolution as it lends itself well to incorporating the 
necessary physics. 
However, alternatives such as the Monte-Carlo method are quickly catching up 
and can cope with larger particle numbers (Fregeau et al. 2003). 
In the meantime, non-direct $N$-body models are very useful for understanding 
the evolution of larger star clusters. 
For a review of star cluster evolution and modeling approaches 
we recommend Meylan \& Heggie (1997). 

In {\tt NBODY4} stellar evolution is provided through the algorithms developed 
by Hurley, Pols \& Tout (2000) which in turn are based on the detailed 
models of Pols et al. (1998). 
As such, the radii and masses of the cluster stars are updated in-step 
with the dynamical evolution of the cluster. 
Binary evolution is included via the rapid binary evolution algorithm 
described in Hurley, Tout \& Pols (2002). 
This covers processes such as steady mass-transfer, common-envelope 
evolution, angular momentum loss through tidal interaction, and coalescence. 
Dynamical encounters between binaries and single stars are followed. 
These can lead to significant perturbation of the orbital parameters of 
binaries and even chaotic orbits or disassociation (Mardling \& Aarseth 2001). 
The formation of triple-star and four-body subsystems, the exchange of 
stars in and out of binary systems and the possibility of collisions 
(hyperbolic or within highly-eccentric binaries) are also accounted for. 
The method in {\tt NBODY4} for modelling the tidal field of the Galaxy is to 
place the cluster on a circular orbit at a 
specified radial distance, $R_{\rm gc}$, from the center of a point-mass, $M_{\rm g}$, 
which is the Galactic mass within $R_{\rm gc}$ (see Giersz \& Heggie 1997).  
As suggested by Chernoff \& Weinberg (1990) the circular speed of the orbit is 
taken as $220 \, {\rm km} \, {\rm s}^{-1}$ for all orbits with $R_{\rm gc} > 3\,$kpc. 
This is clearly an idealized model of the Galaxy and more elaborate external potentials 
will be explored in future models. 
Disk shocking -- the tidal shocks that a cluster receives as it passes through the Galactic disk 
(Chernoff, Kochanek \& Shapiro 1986) -- is an option in {\tt NBODY4} but is not implemented in the 
models presented here. 
A full overview of the {\tt NBODY4} code is given in Aarseth (2003). 

\subsection{Description of the Model Family} 
\label{s:met_desc}

In this work we have a series of {\tt NBODY4} models at our disposal that 
each started with $N = 100\,000$ and were 
performed on the GRAPE-6 boards \citep{mak02} 
located at the American Museum of Natural History. 
The setup of each model was identical except for the fraction of stars in primordial 
binaries and the strength of the Galactic tidal field. 
In terms of primordial binaries the fractions considered were $f_{\rm b} = 0.0$, 0.05 and 0.1 
where $f_{\rm b} = N_{\rm b} / \left( N_{\rm s} + N_{\rm b} \right)$, 
for $N_{\rm s}$ single stars and $N_{\rm b}$ binaries, and $N = N_{\rm s} + N_{\rm b}$. 
For the tidal field, clusters were placed on circular orbits at either 
$R_{\rm gc} = 4\,$kpc or $8.5\,$kpc (the Solar distance) from the Galactic 
center using the point-mass galaxy approach outlined above. 
In summary, the models available for analysis are: 
\begin{itemize} 
\item 0\% primordial binaries and orbiting at $4\,$kpc; 
\item 0\% primordial binaries and orbiting at $8.5\,$kpc;
\item 5\% primordial binaries and orbiting at $8.5\,$kpc; and, 
\item 10\% primordial binaries and orbiting at $8.5\,$kpc. 
\end{itemize}
Two realizations of each model were evolved and results presented will be an 
average of these unless otherwise stated. 

In each starting model the masses of 
the single stars were drawn from the IMF 
of \citet{kro93} between the mass limits of 0.1 and $50 M_\odot$.
Each binary mass was chosen from the IMF of \citet{kro91}, 
as this had not been corrected for the effect of binaries, 
and the component masses were set by choosing a mass-ratio from a uniform distribution. 
This gave an initial total mass of about $52\,000 M_\odot$. 
The orbital separations of the primordial binaries were drawn from the 
log-normal distribution suggested by \citet{egg89} with a peak at $30\,$au. 
A maximum separation of $100\,$au was imposed. 
Orbital eccentricities were assumed to follow a thermal distribution \citep{heg75}. 
We used a Plummer density profile \citep{plu11} and assumed the stars 
are in virial equilibrium when assigning the initial positions and 
velocities. 
The Plummer model is a common choice for the initial density distribution of a model 
star cluster because of its ease of use numerically (e.g. Aarseth, H\'{e}non \& Wielen 1974). 
There is no evidence that such a model approximates the real density profile  
of a globular cluster at birth. 
However, dynamical evolution quickly erases the Plummer profile signature. 

A metallicity of $Z = 0.001$ was assumed for the stars in these simulations. 
This is higher than the actual value for NGC$\,6397$ which is $Z \sim 0.0002$ 
based on $[ {\rm Fe/H} ] = -1.95$ (Harris 1996). 
However, Hurley et al. (2004) have demonstrated that there is little to no difference 
in the evolution of cluster models that switch from $Z = 0.001$ to $0.0001$. 
This is because the stellar evolution and mass-loss timescales are similar 
for these two metallicities. 
Conversely there are significant differences, such as the timescale for core-collapse,  
between models based on Population I ($Z = 0.02$) and Population II 
(typically $Z = 0.001$ or lower)  composition. 
This is a direct result of differing stellar evolution timescales 
-- the main-sequence lifetime of a solar-mass Population II star is a factor of two 
shorter than for its Population I counterpart, for example. 

\subsection{The Main $N$-body Model} 
\label{s:met_main}

It has been established for some time now that globular clusters are born 
with at least some primordial binaries (Hut et al. 1992) so the 0\% models can 
be considered as reference models rather than models for comparison 
to NGC$\,6397$. 
On the other hand, observations of NGC$\,6397$, combined with the results of dynamical models 
that will be discussed further in Section~\ref{s:res_fb}, indicate only a modest 
fraction of primordial binaries (Davis et al. 2008b). 
As such the model with 5\% primordial binaries is preferred to that with 10\%. 

The orbit of NGC$\,6397$ within the Galaxy ranges from a perigalacticon of $\sim 3\,$kpc 
to an apogalacticon of $\sim 7\,$kpc (Dauphole et al. 1996; Milone et al. 2006; Kalirai et al. 2007). 
Baumgardt (2001: see also King 1962) showed that when approximating the orbit of a cluster 
as circular it is preferable to place the 
cluster near its perigalactic distance -- for NGC$\,6397$ that means it would be best 
represented by a model cluster orbiting at $3$--$4\,$kpc from the Galactic centre. 
Unfortunately the model placed at $4\,$kpc has dissolved by an 
age of $9\,$Gyr and is thus of little use in comparison to an old globular cluster 
such as NGC$\,6397$. 
In contrast, the models placed at $8.5\,$kpc have $\sim 50$\% of their stars remaining 
after $9\,$Gyr of evolution and 10\% remaining when the simulations are halted 
at $20\,$Gyr. 
Having a model orbit closer to the Galactic center 
speeds up the cluster evolution and 
hastens cluster dissolution -- effectively the stronger tidal field gives a smaller tidal 
radius which in turn means a smaller half-mass radius and a shorter half-mass 
relaxation timescale. 
That in turn means earlier core-collapse 
-- the $100\,000$ star model orbiting at $4\,$kpc reached core-collapse at $6\,$Gyr 
compared to $\sim 16\,$Gyr for the models orbiting at $8.5\,$kpc. 

It is certainly important for our investigation of NGC$\,6397$ to compare the observed 
data with a model that is at a similar dynamical age. 
That means post-core collapse. 
However, we must be careful not to extend the comparison too far into this phase 
as mass-segregation and the escape of stars, for example, will continue to develop 
and affect the model parameters. 
Notably, Baumgardt \& Makino (2003) have demonstrated that the modification of the 
MF slope by dynamical evolution can be quantified in terms of the fractional mass lost 
from the cluster (see their eq.~14). 
Our $N = 100\,000$ model at $8.5\,$kpc has 20\% of its initial mass remaining at $16\,$Gyr. 
Using  eq.~11 of Vesperini \& Heggie (1997), with the escape rate of stars and amount of 
stellar evolution mass-loss calibrated to this model, we can estimate the starting $N$ 
required to give the same fractional mass but at an age of $12\,$Gyr 
for a cluster orbiting at $4\,$kpc (the approximate age and $R_{\rm gc}$ of NGC$\,6397$). 
The answer is $N \sim 200\,000$. 
We can also estimate, based on the initial half-mass relaxation timescale, that such a 
model would reach core collapse at $\sim 11\,$Gyr. 
This would therefore be a more appropriate model for comparison to NGC$\,6397$ 
-- we have started a model with these parameters but it will take at least a year to complete. 
As discussed in Section~\ref{s:disc} an even larger starting $N$ than $200\,000$
is required in order to generate a direct model of NGC$\,6397$ and such an $N$ 
is not feasible with current computational capabilities. 
We emphasise at this early stage that the models used in our current study are not 
direct models of NGC$\,6397$. 
The impact of this on our analysis as well as 
future improvements to the model, including consideration of a more elaborate 
tidal field with tidal shocks, will also be discussed further in Section~\ref{s:disc}. 

All things considered, the model starting with $N = 100\,000$, 5\% primordial binaries, and 
on a circular orbit at $8.5\,$kpc from the Galactic center will be the focus of our study. 
This is our main model for comparison to the NGC$\,6397$ data and as such 
will often be referred to simply as {\it the} $N$-body model. 
The remaining models will be used for comparative purposes. 

In using our main model we will be applying the results of 
post-core-collapse models of $16\,$Gyr or more to understand 
observations of a post-core-collapse cluster that is $12\,$Gyr old.  
So we have a good model in terms of the dynamical state but the 
stellar evolution ages of the simulated and actual clusters 
are out of synch by at least $4\,$Gyr. 
This is not so much of a concern as it may seem at first glance because 
stellar evolution, and in particular mass-loss from stellar evolution, 
is most important during the early stages of cluster evolution and its role 
diminishes as the cluster population ages 
(unless the cluster is close to complete dissolution). 
Figure~\ref{f:fig2} compares the mass-loss from stellar evolution 
to that of escaping stars as a function of cluster age for our main model. 
We can clearly see that the stellar evolution mass-loss rate slows down with time 
and that escaping stars  dominate in the latter stages. 
As noted by Baumgardt \& Makino (2003), the bulk of mass-loss owing to stellar evolution 
is complete prior to other, dynamically-driven, mass-loss mechanisms coming into play. 
Consider also that the turn-off mass for a $Z = 0.001$ cluster only decreases from $0.83$ 
to $0.77 M_\odot$ between $12$--$16\,$Gyr and that the mass of the white dwarfs being 
produced only varies by $0.02 M_\odot$ over the same timeframe. 
So the effect of the mismatched stellar evolution age on the mass-segregation observed 
in the post-core-collapse models is expected to be negligible.

\subsection{Evolution Overview} 

To conclude this section we will give an overview of the evolution 
of the main $N$-body model. 
The initial parameters of the $95\,000$ single stars and $5\,000$ binaries 
were determined as described above in Section~\ref{s:met_desc}. 
Residing at $R_{\rm gc} = 8.5\,$kpc and with an initial mass of $51\,910 M_\odot$ 
produced a tidal radius, $r_{\rm t}$, of about $50\,$pc for the initial model. 
The initial half-mass radius, $r_{\rm h}$, was $6.7\,$pc. 
Note that as demonstrated by Hurley (2007) the two-dimensional projected 
half-light radius is typically a factor of two less than the three-dimensional 
half-mass radius. 
Formally, the Plummer model used for the initial density profile extends to infinite 
radius but in practice stars placed at great distances are not accepted for the starting 
model. 
The ratio of the Plummer model scale-radius, $b$, to the tidal radius of the 
initial model was 0.1 while the ratio of the radius containing the inner 99\% of 
the cluster mass, $r_{\rm 99}$, to $r_{\rm t}$ was 0.77 (0.02\% of the cluster 
by mass lay outside of $r_{\rm t}$ initially). 
We note that it takes of order a half-mass relaxation time for the cluster to evolve 
to a point where the density profile starts to resemble that of a real globular 
cluster (as described by King 1962). 
As such it is relevant to compare radii of the model at an age of $2.2\,$Gyr 
(when the evolution age and half-mass relaxation timescale were equal) 
to the tidal radius: 
$r_{\rm 99}/r_{\rm t} = 0.72$ and $r_{\rm h}/r_{\rm t} = 0.19$ (increased from 0.13 initially). 

\begin{figure}
\plotone{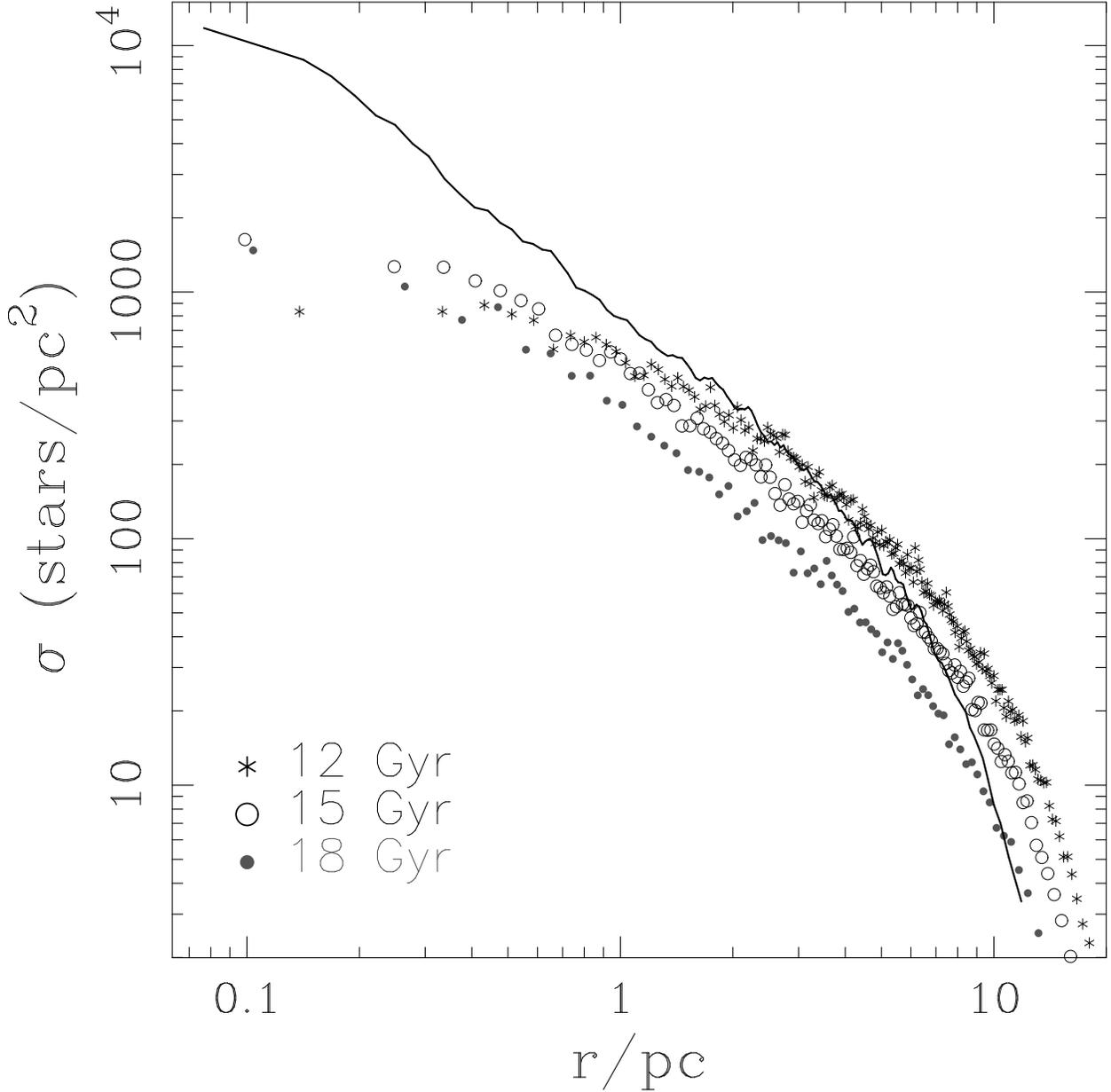}
\caption{
The projected surface density profiles for the model cluster at ages of 12, 15 and $18\,$Gyr. 
Data are from the simulation starting with $N = 100\,000$, 5\% binaries and 
orbiting at $8.5\,$kpc from the Galactic center. 
All stars bound to the cluster at the specified age are included and the profile is 
constructed with 200 stars per bin. 
Note that smoother profiles can be achieved with larger bins but that this would remove 
the `cuspy' features in the central regions. 
Conversely, it is possible to probe deeper into the core with smaller bin sizes but 
the data points then become too noisy. 
Note also that the profiles have not been normalized to the total number of stars 
remaining in the cluster at the given age -- if they were then it would be clear 
that the $18\,$Gyr profile has the steepest slope of the three profiles shown for this model. 
The solid line shows the profile from the simulation starting with $100\,000$ stars, 
0\% binaries and orbiting at $4.0\,$kpc from the Galactic center, at an age of $5.6\,$Gyr.  
\label{f:fig1}}
\end{figure}

\begin{figure}
\plotone{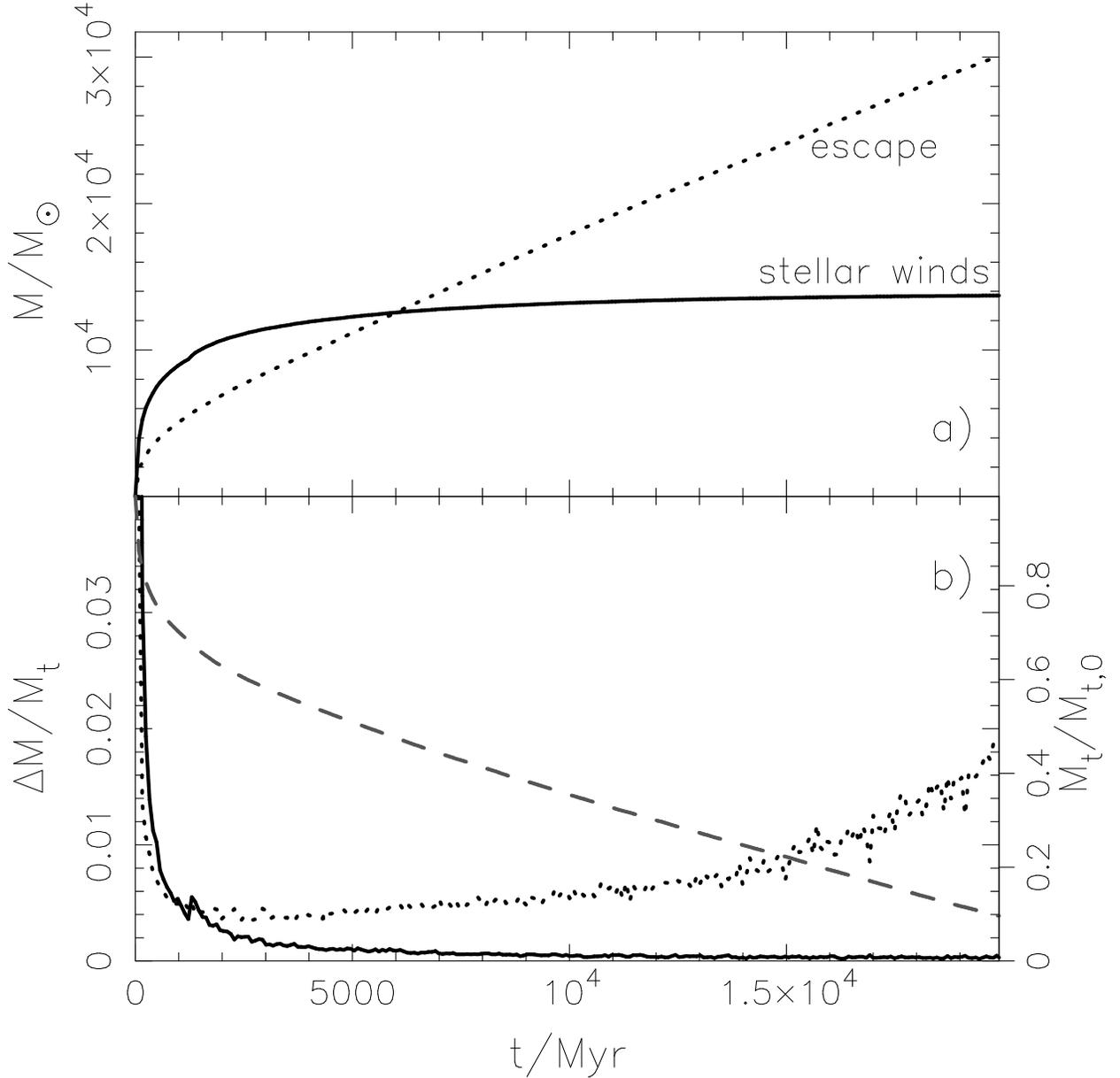}
\caption{
Mass-loss from the model cluster as a function of age. 
The upper panel shows the cumulative mass-loss from stellar winds (solid line) 
and from the escape of single stars (dotted line). 
The lower panel shows the mass lost per snapshot interval ($100\,$Myr in this simulation) 
scaled by the average total mass, $M_{\rm t}$, of the cluster in that interval 
-- this demonstrates the relative mass loss rates from stellar winds (solid line) 
and stellar escape (dotted line). 
The sharp increase in the stellar wind mass-loss rate at $\sim 1.3\,$Gyr is real and 
corresponds to the turn-off mass dropping below $1.8 M_\odot$ which is the maximum 
mass for degenerate helium ignition at the tip of the giant branch (for a $Z = 0.001$ 
population). 
Also shown in the lower panel is the evolution of the total mass of the cluster 
(scaled by the initial total mass: grey dashed line) for reference purposes. 
This decreases from $1.0$ at $0\,$Gyr to $0.09$ at $20\,$Gyr (as per the axis label 
on the right-side of the panel). 
\label{f:fig2}}
\end{figure}

The model cluster was evolved to an age of $20\,$Gyr. 
At regular intervals ($100\,$Myr) snapshots of the cluster are saved 
which enable post-simulation analysis of the cluster structure and stellar populations. 
Events occurring between snapshots, such as collisions, are also stored. 
After an initial phase of expansion the size of the cluster core steadily 
decreased until reaching a minimum at about $16\,$Gyr of age. 
From that point on the cluster is assumed to be post-core-collapse. 
The half-mass relaxation timescale at core-collapse was $500\,$Myr, 
reduced from $1\,400\,$Myr initially. 
The core density increased from $100 \, {\rm stars} \, {\rm pc}^{-3}$ initially to 
a local maximum of $2 \times 10^3 \, {\rm stars} \, {\rm pc}^{-3}$ at $16\,$Gyr 
and a maximum of $10^4 \, {\rm stars} \, {\rm pc}^{-3}$ just prior to termination 
at $20\,$Gyr. 
The cluster parameters at various times in the evolution are summarised 
in Table~\ref{t:table1}. 
In Figure~\ref{f:fig1} we show the evolution of the cluster density profile. 
This demonstrates that the profile steepens as the model evolves from 
pre-core-collapse ($12\,$Gyr), to near core-collapse ($15\,$Gyr) and then 
post-core-collapse ($18\,$Gyr). 
For comparison, the profile of the cluster evolved on an orbit with $R_{\rm gc} = 4\,$kpc 
is shown at an age of $5.6\,$Gyr. 
This is just prior to the end of the core-collapse phase and finds the model 
cluster at a comparable dynamical age to that of our main model at $15\,$Gyr. 
Clearly the stronger tidal field experienced by the model orbiting closer to the 
Galactic center has lead to 
a more centrally-concentrated cluster.

\section{Results}
\label{s:result}

We now look at results of the $N$-body model that are particularly relevant to 
analysis of the NGC$\,6397$ observations. 
All results refer to the main $N$-body model (see Section~\ref{s:met_main}) 
unless otherwise stated. 
At times we will talk of the `outer' region of the cluster which is intended to 
correspond to the ACS `outer' field of NGC$\,6397$. 
The observed field covers a region approximately $2$--$3$ half-light radii from 
the cluster center which for the model corresponds to $1$--$1.5$ half-mass radii 
from the model center. 
As such, `outer' is only used in comparison to a central, or `inner', region 
and does not refer to the very outer, or halo, of the cluster which extends to 
the tidal radius (typically $5$--$6$ half-mass radii).

\subsection{Colour-Magnitude Diagram}
\label{s:res_cmd}

\begin{figure}
\plotone{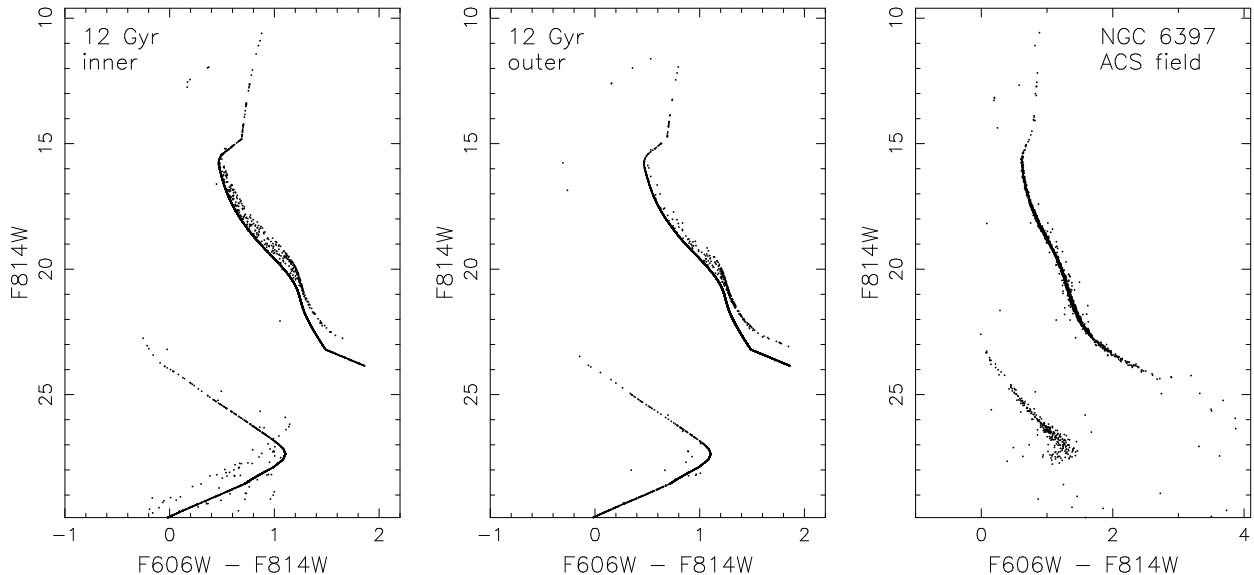}
\caption{
Color-magnitude diagrams for the $N$-body model at $12\,$Gyr. 
The left panel shows the CMD for the central/inner region of the model cluster and includes stars out to 
$3\,$pc (the half-light radius or about 3 core radii). 
The middle panel is for a region between two and three half-light radii from the cluster center 
(similar to the ACS field). 
Both model regions contain approximately $6\,500$ objects. 
Binaries are assumed to be unresolved. 
The right panel shows the proper motion cleaned CMD of stars observed in the ACS 
field of NGC$\,6397$ (data taken from Fig. 5 of Richer et al. 2007). 
For the $N$-body models conversion to ACS colors utilizes the model atmosphere calculations 
of Castelli \& Kurucz (2003), except for white dwarfs where we use the 
results of Bergeron, Wesemael \& Beauchamp (1995). 
A distance modulus of $12.36$ is assumed. 
Note that only one of the $N$-body models starting with $N = 100\,000$ and 5\% binaries 
is used to create the model CMDs. 
\label{f:fig3}}
\end{figure}

In Figure~\ref{f:fig3} we show the model color-magnitude diagram (CMD) at $12\,$Gyr for stars in 
the region corresponding to the ACS field (the outer field) and compare this 
to the CMD of stars in the center of the model. 
We have not attempted to synthesize photometric error in order to produce scatter in 
our model CMDs. 
As such, the narrow, well-defined sequences in the model CMDs are composed of 
single stars while the more scattered sequences are produced primarily by 
unresolved binaries. 
It is evident that the inner region is more abundant in giant and horizontal branch stars, 
as we would expect from mass-segregation, 
while the outer region is almost devoid of binaries on the upper main-sequence (MS). 
We note that the lower end of the MS in Figure~\ref{f:fig3} extends only to $F606W - F814W \sim 2$ 
while the observed MS of NGC$\,6397$ reaches $F606W - F814W \sim 4$ at the 
hydrogen-burning limit (see the right-hand panel of Figure~\ref{f:fig3} and 
also Fig.~6 of Richer et al. 2007). 
This is simply a result of choosing $0.1 M_\odot$ as the lower limit to the IMF used in 
the $N$-body models whereas the mass at the hydrogen-burning limit is reported 
to be $0.083 M_\odot$ (Richer et al. 2007). 
The difference in $F606W - F814W$ between models of $0.1$ and $0.083 M_\odot$ 
(taken from the stellar models of Dotter et al. 2007) corresponds roughly to the difference 
between the end-points of the model and observed CMDs. 
Note also that reddening has not been applied to the model CMD but this gives 
only $E \left( F606W - F814W \right) \sim 0.2$ (Hansen et al. 2007). 

An age of $12\,$Gyr corresponds to the approximate stellar evolution age of NGC$\,6397$. 
However, as discussed in Section~\ref{s:met_main}, it is more pertinent for our study to 
investigate models that correspond to the dynamical age of the cluster. 
With this in mind we show in Figure~\ref{f:fig4} the inner and outer field CMDs at a 
model age of $16\,$Gyr, when the model cluster has reached the post-core-collapse phase. 
Once again giant stars, horizontal branch stars and upper-MS binaries are more 
abundant in the inner field compared to the outer field. 
We now also have blue straggler stars present but residing in the inner region only. 
This is a clear demonstration of the `binary-burning' effect expected in globular cluster 
cores during the late stages of core-collapse (McMillan, Hut \& Makino 1990). 
We also see that in the outer region the white dwarf (WD) sequence is very `clean', 
meaning that there is 
minimal contamination from double-white dwarf binaries and ÔdivorcedÕ single white dwarfs 
(those that were members of a binary and have followed a non-standard evolution path as a 
result of mass-transfer and/or a merger). 
Note also that hot WDs are more likely to be found in the inner field than the outer field. 
These points regarding white dwarfs will be discussed further in Section~\ref{s:res_wd}. 
For comparison we have reproduced the NGC$\,6397$ CMD obtained by Richer et al. 
(2007: see their Figure~5) in both Figures~\ref{f:fig3} and \ref{f:fig4}. 
We find that in all of the obvious stellar population aspects that the outer field CMD of the 
model at $16\,$Gyr compares well to the corresponding 
NGC$\,6397$ field (see Richer et al. 2007).

\subsection{Binary Fractions}
\label{s:res_fb}

Davis et al. (2008b) have derived a binary fraction of 0.02 for the ACS outer field of NGC$\,6397$. 
As shown by Hurley, Aarseth \& Shara (2007) this should be a good indicator of the 
primordial binary fraction of the cluster 
-- owing to mass-segregation the binary fraction in the core increases with time 
but outside of the half-mass radius the binary fraction varies little from the primordial value. 
To emphasize this point we show in Figure~\ref{f:fig5} the evolution with cluster age of the binary 
fractions in the inner and outer fields of our model. 
After $16\,$Gyr the inner binary frequency is 9\% and the outer frequency is 5\% (same as 
the primordial binary value). 
This is when all cluster stars (and binaries) are considered. 
The binary fraction derived by Davis et al. (2008b) was for MS objects only and we show 
the corresponding binary fraction in Figure~\ref{f:fig5}. 
In the center of the cluster this fraction is noticeably higher (0.15 at $16\,$Gyr) 
than the overall fraction. 
However, in the outer field there is not such a clear distinction and the MS binary fraction 
is similar to the primordial binary fraction. 

For interest sake we also show in Figure~\ref{f:fig5} the MS binary fraction if MS-MS 
binaries with mass-ratios, $q$, less than 0.5 are incorrectly classified as single stars. 
This is to mimic observational methods that identify binaries through inspection of the 
CMD MS and cannot distinguish low-$q$ binaries from single MS stars as they do not 
separate clearly from the single MS. 
This gives an underestimate of the true binary fraction. 
However, Figure~\ref{f:fig5} shows that the error induced is not large (although the 
effect is dependent on the assumed $q$ distribution of the primordial binaries). 
The choice of $q = 0.5$ as the boundary mass-ratio was partly motivated by the 
work of Hurley \& Tout (1998) which showed that binaries with $q = 0.5$, and 
primary masses in the range $1$--$2 M_\odot$, sit very close to the single star MS. 
Davis et al. (2008b) have shown that this condition could be relaxed to $q = 0.4$, 
or even 0.3, for the portion of the CMD that is of interest for the analysis of the 
NGC$\,6397$ data 
($F814W > 17$, corresponding to primary masses of $0.75 M_\odot$ or less). 
If we instead classify binaries with $q < 0.3$ as single stars the difference between 
the derived binary fraction and the correct MS binary fraction 
is greatly reduced (compared to using $q < 0.5$). 
Of course, when deriving cluster binary fractions the existence of low-$q$ binaries 
amongst the single star MS can be modeled by making assumptions about the 
underlying $q$ distribution, as has been done by Davis et al. (2008b). 
What is more difficult to account for is the possible contamination owing to MS-WD 
binaries where the WD is sufficiently cool that the MS star dominates the light. 
Such binaries have been counted as single stars when calculating the MS binary 
fractions shown in Figure~\ref{f:fig5} but do not make a significant contribution 
(2\% or less of the stars).

\subsection{Mass Functions}
\label{s:res_mf}

In Figure~\ref{f:fig6} we look at how the mass function varies with radius 
in a model just prior to the end of the core-collapse phase ($15\,$Gyr). 
This is supplemented by Figure~\ref{f:fig7} which does the same for a 
post-core-collapse model ($18\,$Gyr). 
The latter age corresponds to the dynamical age assumed by Richer et al. (2007) 
in their analysis of the NGC$\,6397$ luminosity function. 
In Figures~\ref{f:fig6} and \ref{f:fig7} 
only MS stars below the cluster turn-off are included as these have suffered no mass-loss 
and are therefore potential indicators of the IMF. 
We divide the cluster into six radial regions chosen such that the number of stars 
in each region is similar. 
For each region we show the local MF and compare this to the global MF of the 
model cluster at that age and the IMF (both are normalized to the local MF). 
Comparison of the global MF to the IMF (see any panel of Figure~\ref{f:fig6}) 
shows clearly the effect of the cluster environment in modifying the appearance of the MF 
-- the evolved MF is deficient in low-mass stars and has an overabundance of high-mass stars. 
This is accentuated in Figure~\ref{f:fig7}, demonstrating that it is an evolutionary effect. 
Basically, mass-segregation causes stars more massive than the average to sink towards the 
center of the cluster while stars of lower mass move outwards. 
As a result low-mass stars are preferentially stripped from the cluster by the tidal 
field of the host galaxy. 
This means that the global MF of a dynamically evolved star cluster can not be used 
to infer the IMF, unless of course the intervening dynamical evolution is properly 
accounted for (as in Richer et al. 2007). 

The process of mass-segregation is clearly evident in the radial variation of the MF 
-- in the center of the cluster massive MS stars are more abundant than low-mass stars 
and the situation reverses as we look progressively further out from the center. 
At both $15$ and $18\,$Gyr the model indicates that the MF in the $1$--$1.5 r_{\rm h}$ region 
(corresponding to the NGC$\,6397$ field) 
gives the best representation of the global MF for the same age. 
This agrees with the earlier findings of King (1996) in relation to NGC$\,6397$. 
In terms of recovering the IMF, a region slightly further out offers the best hope 
at $15\,$Gyr ($2$--$3 \, r_{\rm h}$) while at $18\,$Gyr one must look even further from the center. 
In Figure~\ref{f:fig8}a we show how this optimal region for recovering the IMF varies 
with cluster age. 
We see that it moves progressively outwards as the cluster evolves and also becomes 
less well defined. 
This result is compared to the behaviour of the model with 0\% primordial binaries 
and orbiting at $4\,$kpc from the Galactic Center in Figure~\ref{f:fig8}b 
(with time scaled by the cluster dissolution time -- see caption for details). 
The scaled behaviour of the models is similar except perhaps at intermediate ages 
-- leading up towards core-collapse -- where the optimal MF region is relatively closer 
to the cluster center for the model cluster on a tighter orbit. 

Davis et al. (2008b) present the MFs for NGC$\,6397$ stars observed in the WFPC2 (central) 
and ACS (outer) fields. 
The MF for MS stars in the central field rises towards higher masses and then 
flattens out 
-- this agrees well with what we observe in the central region of the model 
(upper-left panel of Fig.~\ref{f:fig6}). 
The MF in the outer field does the opposite, rising towards lower masses, which 
again is in good agreement with the corresponding model region 
(middle-left panel of Fig.~\ref{f:fig6}). 
However, Davis et al. (2008b) find a turnover in the outer field MF at a mass of about $0.2 M_\odot$ 
which is not observed in the model MF. 
The MF of the model cluster in the $1.0 < r/r_{\rm h} < 1.5$ region certainly does flatten at 
the low-mass end as the cluster evolves (comparing Figs.~\ref{f:fig6} and \ref{f:fig7}) but at 
no stage do we see a turn-over at the low-mass end, even when looking at the very 
outer regions. 
One could expect that this behaviour will be dependent on the strength of the external tidal field 
-- a stronger tidal field will be more efficient at removing low-mass stars. 
Taking our $100\,000$ star model evolved on an orbit at $4\,$kpc from the Galactic Center 
(as opposed to $8.5\,$kpc) and looking at the MFs at an age of $7\,$Gyr we do see evidence 
of the MF `turning-over' at the low-mass end. 
We note though that the data are rather noisy in this case and we will need to wait until 
simulations with a larger initial $N$ are completed before investigating this fully.

\subsection{White Dwarfs}
\label{s:res_wd}

\begin{figure}
\plotone{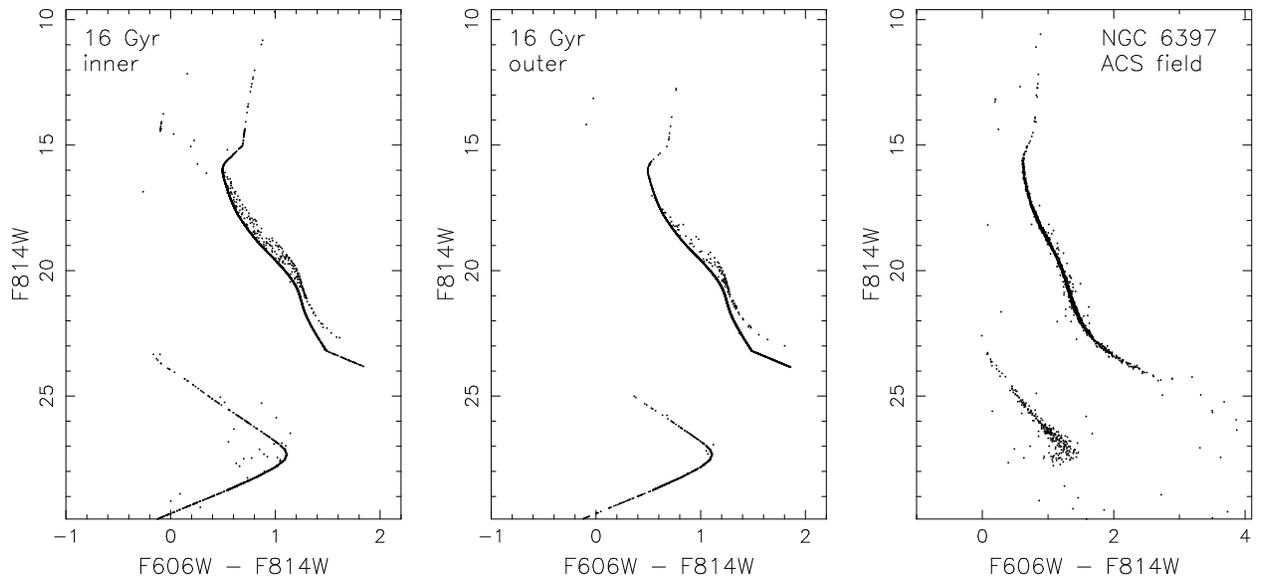}
\caption{
As for Figure~\ref{f:fig3} but at $16\,$Gyr. 
The inner region (left panel) extends out to 
$2.4\,$pc (the half-light radius or about 14 core radii). 
The outer region (middle panel) includes stars between two and three half-light radii 
from the cluster center.  
Both model regions contain approximately $4\,500$ objects. 
Once again, for comparison, the proper motion cleaned NGC$\,6397$ CMD is 
also shown (right panel). 
\label{f:fig4}}
\end{figure}

\begin{figure}
\plotone{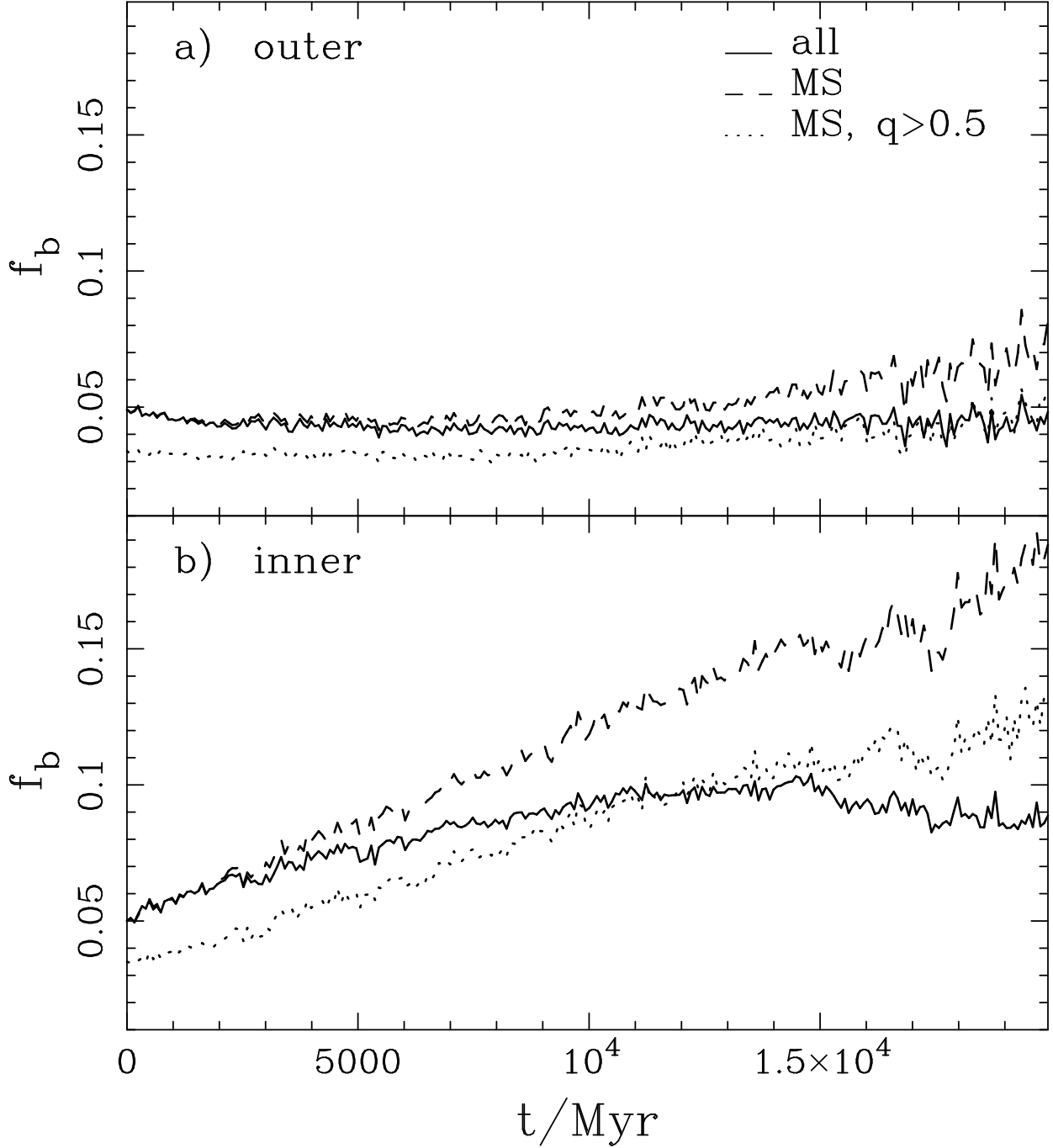}
\caption{
Binary fraction as a function of cluster age: 
a) the outer region ($1$--$1.5 \, r_{\rm h}$); and, 
b) the central/inner region ($0$--$0.5 \, r_{\rm h}$) of the cluster. 
The solid lines give the binary fraction based on all cluster stars while the dashed lines are 
for stars and binaries on the main-sequence only. 
Included in the latter are MS-WD binaries where the WD has cooled sufficiently that the 
unresolved binary lies on the MS. 
The dotted lines are the MS binary fractions but with low-mass ratio binaries ($q < 0.5$) 
counted as single stars. 
\label{f:fig5}}
\end{figure}

\begin{figure}
\plotone{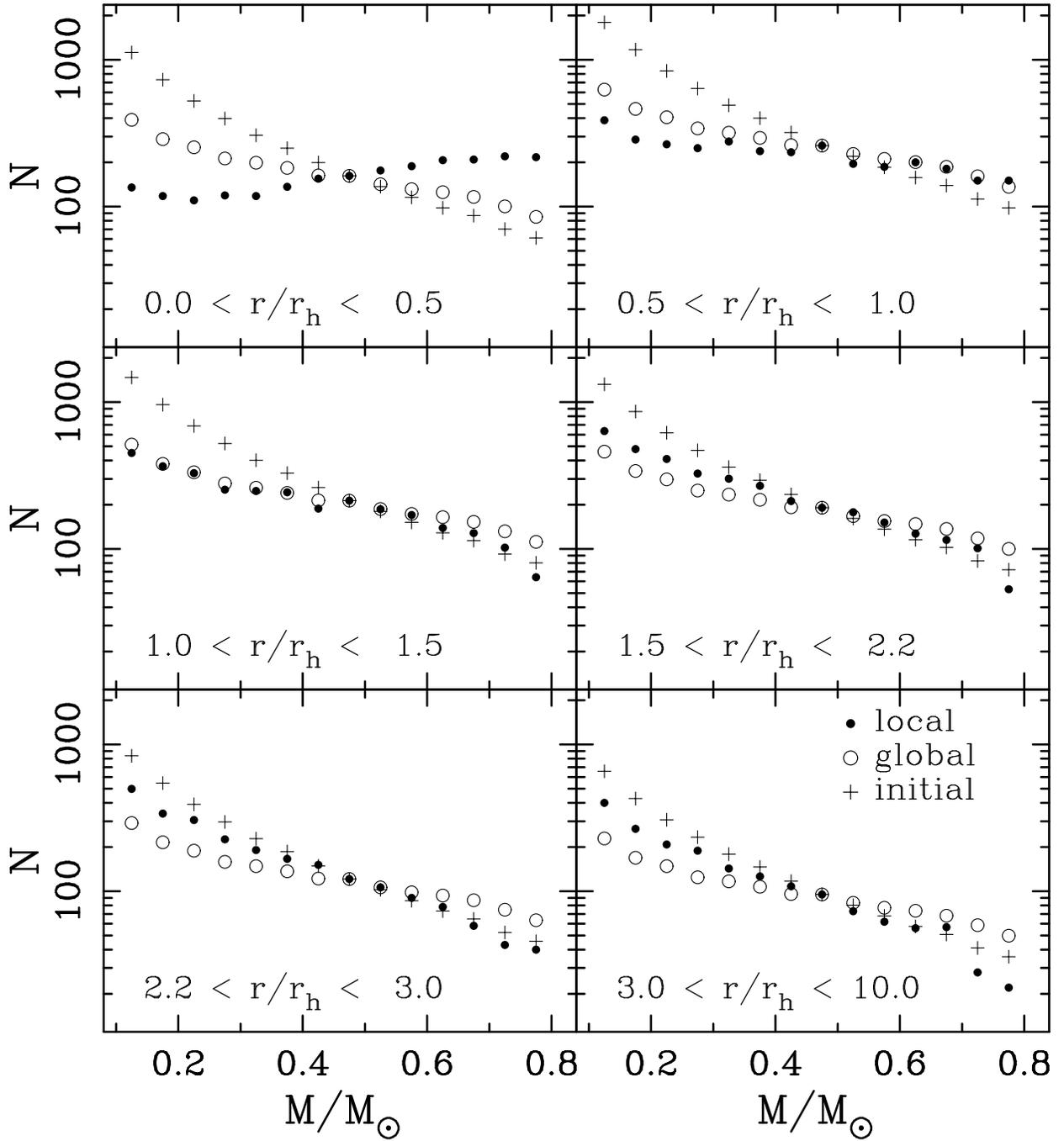}
\caption{
Mass functions of main-sequence stars in various regions of the model cluster at $15\,$Gyr 
(moving outwards as we move from left to right and top to bottom). 
In each region the local MF (solid circles) is compared to the normalized global MF 
(open circles) and also the IMF (+ symbols).  
The regions were chosen to give a similar number of stars in each. 
\label{f:fig6}}
\end{figure}

\begin{figure}
\plotone{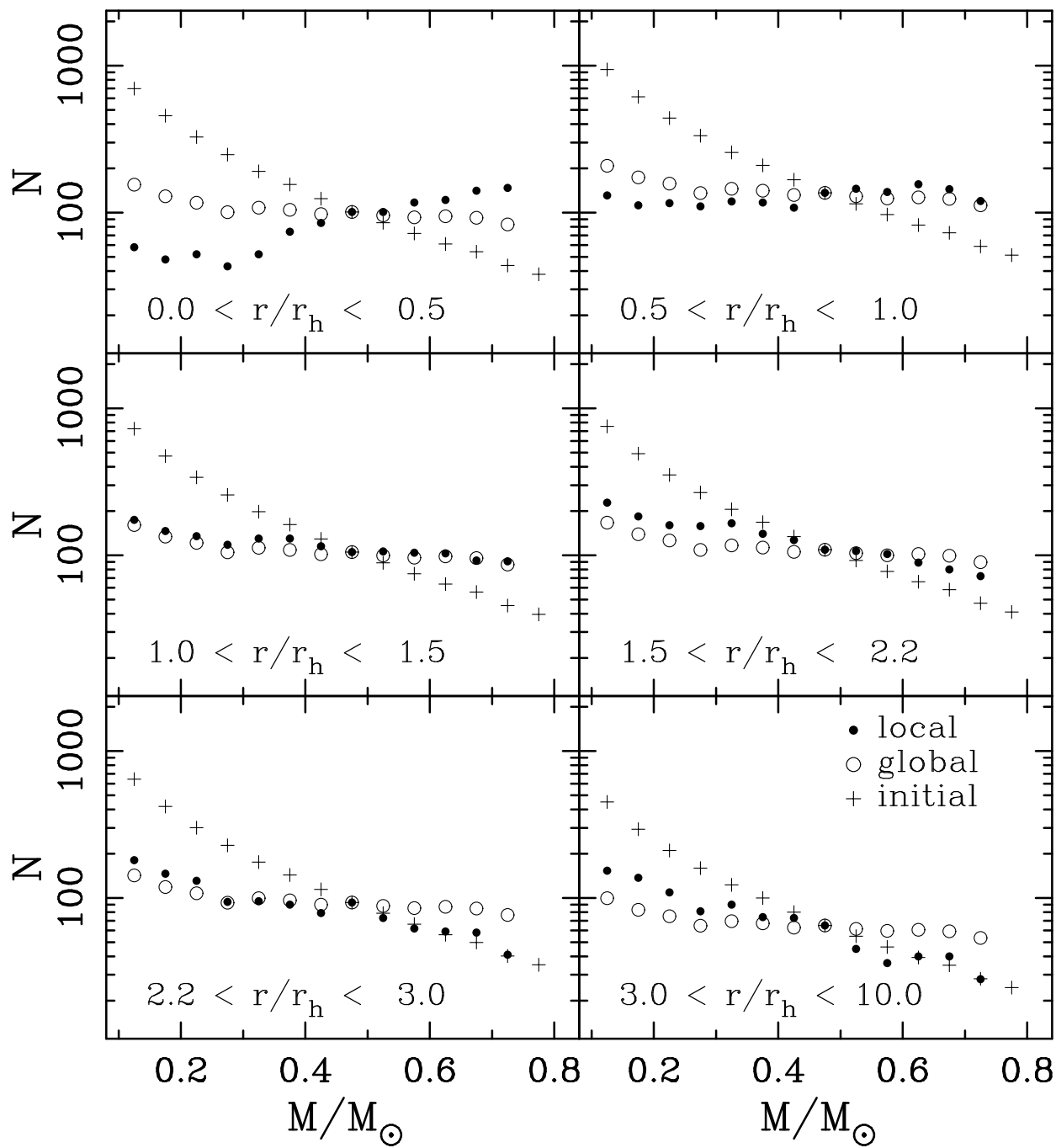}
\caption{
As for Figure~\ref{f:fig6} but at $18\,$Gyr. 
\label{f:fig7}}
\end{figure}

\begin{figure}
\plotone{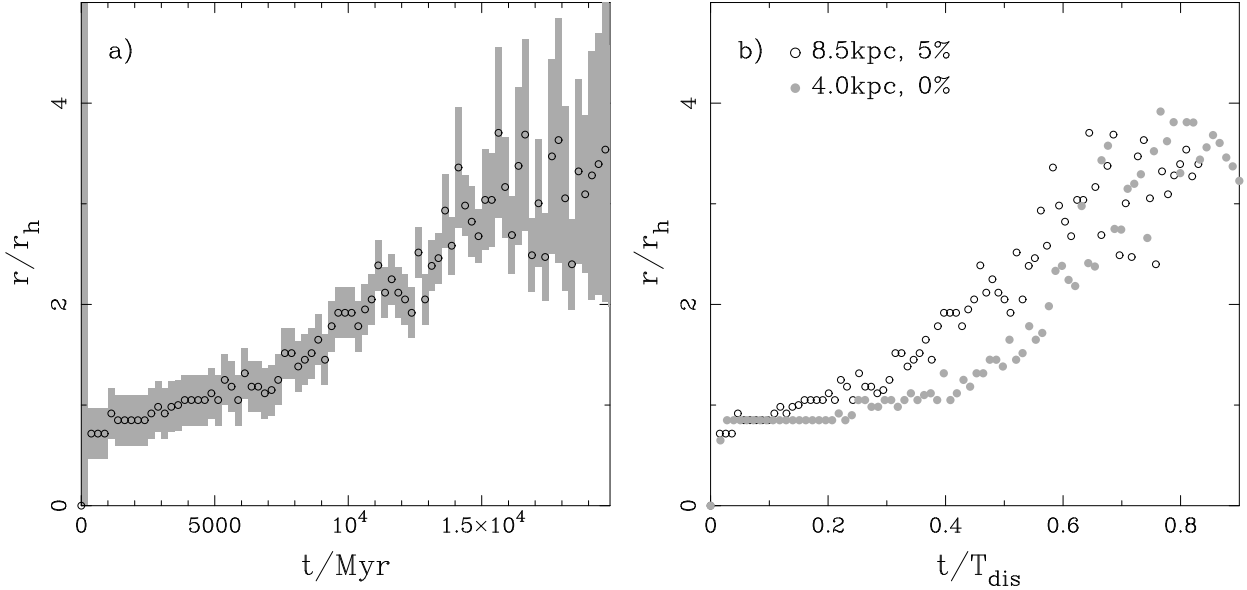}
\caption{
The radial position in the cluster (scaled by the half-mass radius) where, 
at a particular age, the MS stars in that region give the best representation of the IMF. 
At each age we start from the center of the model cluster with a region $0.5 r_{\rm h}$ 
in width and move this region outwards in radius in $0.2 r_{\rm h}$ increments, 
forming the local MF and doing a least-squares fit between this and the IMF each time 
the region is moved. 
The left panel shows the result for our main $N$-body model (open circles) where the 
region with the minimum least-squares statistic is covered by the grey-shaded area 
(expanded if adjoining regions give a least-squares statistic within 10\% of the minimum). 
In the right panel the result is compared to our model evolved at $4\,$kpc from the 
Galactic Center (grey-scale solid circles) which began with 0\% primordial binaries 
as opposed to 5\% for the main model. 
To provide a meaningful comparison the time is scaled by the dissolution time of 
each model, $T_{\rm dis}$, which we take as the time when only 1\% of the original 
cluster remains ($24.2\,$Gyr for the main model orbiting at $8.5\,$kpc and $8.9\,$Gyr 
for the model orbiting at $4\,$kpc). 
\label{f:fig8}}
\end{figure}

\begin{figure}
\plotone{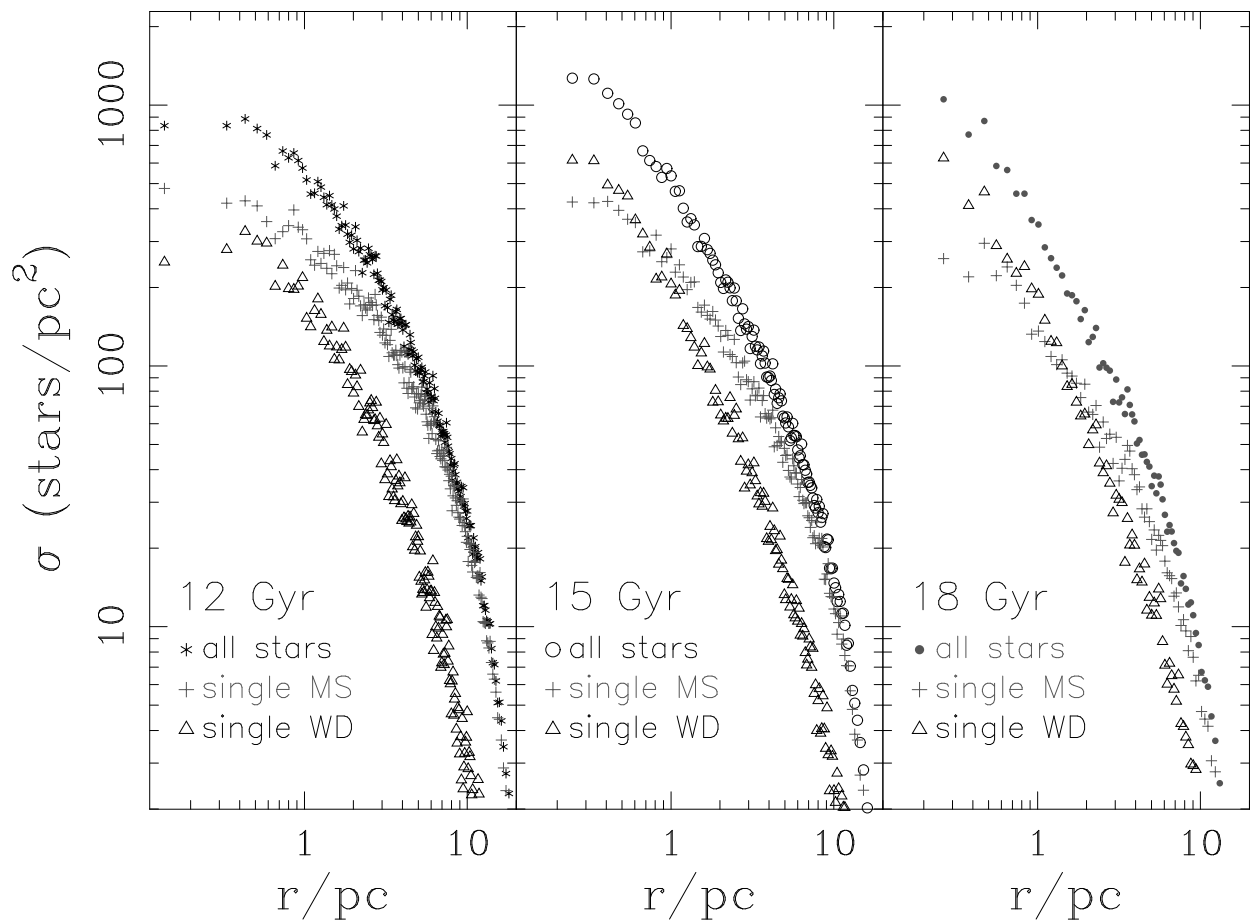}
\caption{
Similar to Figure~\ref{f:fig1} but now also looking at the density profiles of the 
single MS stars (grey-scale + symbols) and single WDs (triangles) at the 
given model ages. 
Note that the profiles for all stars correspond exactly to those shown in Figure~\ref{f:fig1}. 
\label{f:fig9}}
\end{figure}

\begin{figure}
\plotone{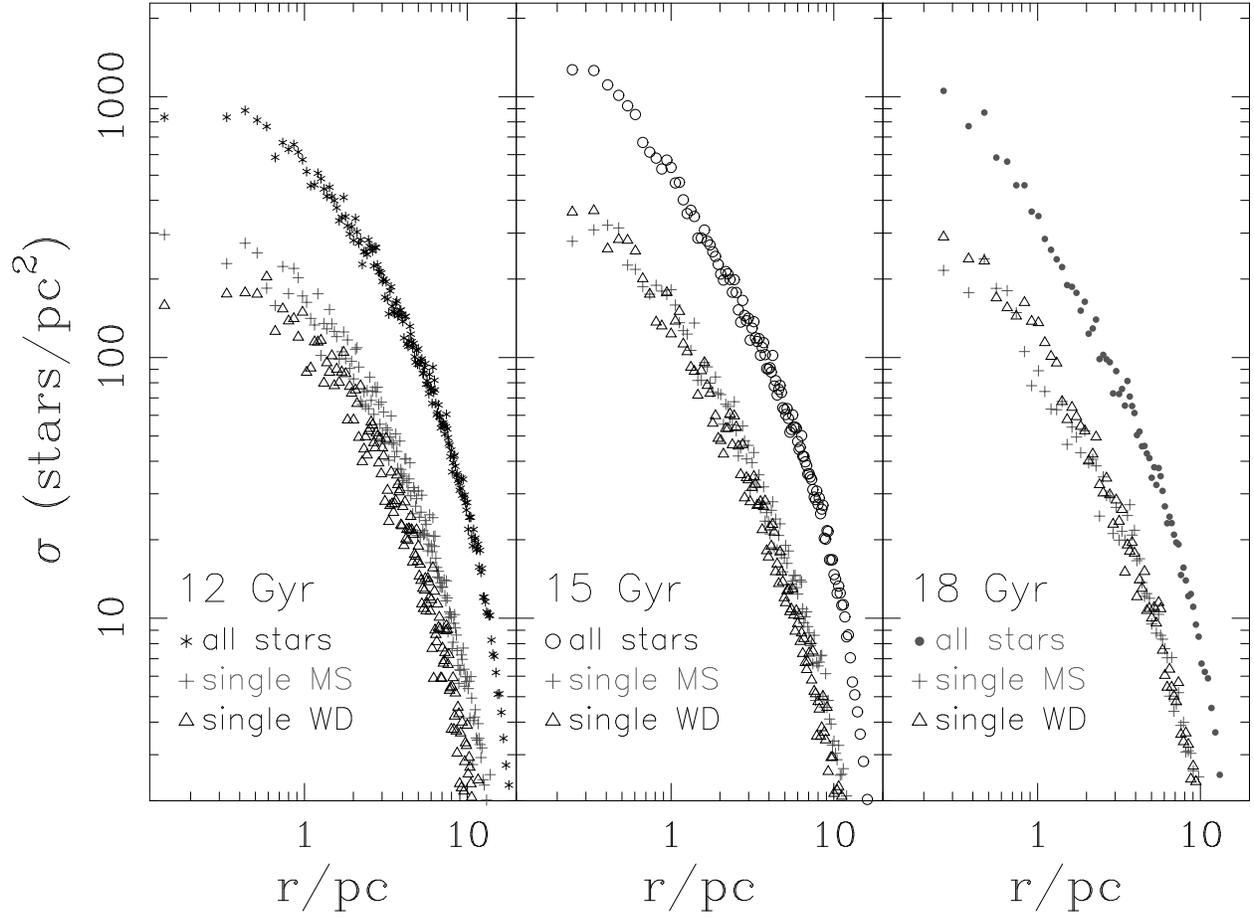}
\caption{
As for  Figure~\ref{f:fig9} but with MS and WD stars restricted to lie 
between $0.5$ and $0.8 M_\odot$. 
\vspace*{0.7cm}
\label{f:fig10}}
\end{figure}

We next turn our attention to the white dwarf population. 
Considering that white dwarfs are born from relatively massive cluster stars 
(which are more likely to reside in the cluster center: see Figure~\ref{f:fig4} or Figure~\ref{f:fig6}) 
it is to be expected that the 
WD population of a dynamically old cluster should be more centrally condensed than that of the 
MS stars. 
This is demonstrated in Figure~\ref{f:fig9}. 
However, it is also true in old clusters that WDs born in the core will segregate outwards in 
radius as they will be less massive than the average mass of objects (stars and binaries) in the core. 
If we instead look at WDs and MS stars in the same mass range 
($0.5$--$0.8 \, M_\odot$: Figure~\ref{f:fig10}) we actually find that they follow similar distributions. 
The point here is that the structure of the WD population will vary across the cluster and the 
variation depends on the stellar and dynamical age of the cluster. 
Another issue that has been raised recently is the possibility that WDs receive a small velocity 
kick at birth (Fellhauer et al. 2003; Davis et al. 2008a). 
The young WDs observed in NGC$\,6397$ have been shown by Davis et al. (2008a) to 
have a more extended radial distribution than predicted by the $N$-body models 
presented here and these models do not include natal kicks for WDs. 

\begin{figure}
\plotone{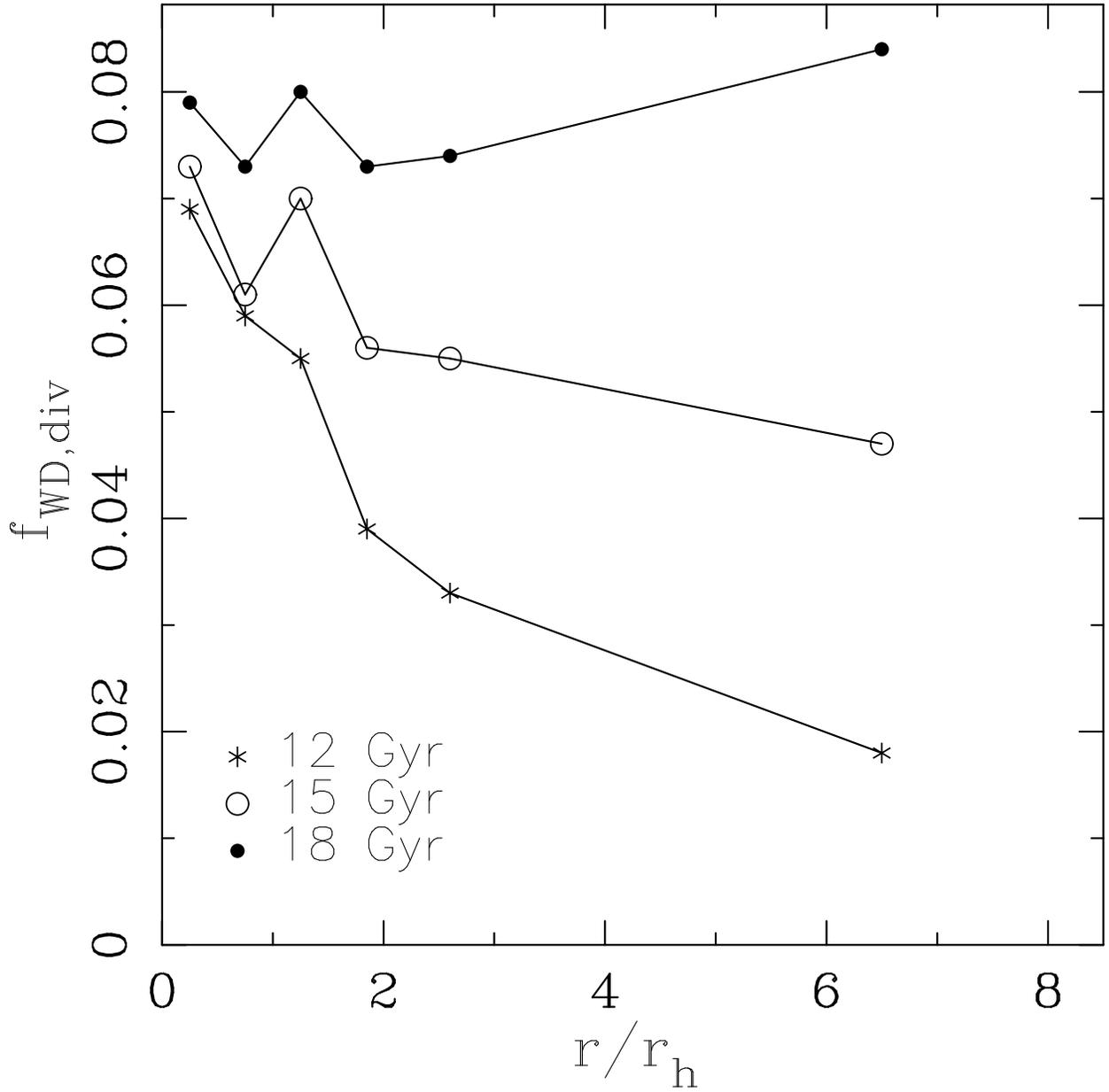}
\caption{
Divorced WD contamination of the WD cooling sequence as a function of radial 
position for ages of $12$, $15$ and $18\,$Gyr. 
Shown on the vertical axis is the ratio of so-called `divorced' single WDs to 
all single WDs (`divorced' + `bachelor': see Hansen et al. 2007). 
\label{f:fig11}}
\end{figure}

Of interest is the possible presence in the WD cooling sequence of double-WD binaries 
and non-standard, or `divorced', single WDs that may affect the determination of the 
true WD cooling age of the cluster. 
Double-WDs are less of a concern as they are generally distinguishable in the CMD. 
They are also less likely to be present in a field outside of the cluster center, 
as is shown in Figure~\ref{f:fig4}. 
However, `divorced' WDs -- those that have had their stellar clock reset by mass-transfer 
in a binary that was subsequently disrupted -- are more difficult to distinguish. 
In Figure~\ref{f:fig11} we show the radial distribution of the divorced-WD fraction at 
ages of $12$, $15$ and $18\,$Gyr. 
Prior to core-collapse the fraction is clearly higher in the center compared to the outer regions 
and it would make sense to observe a field in the latter. 
As the cluster becomes further evolved the fraction of divorced-WDs increases and 
the radial variation diminishes. 
However, the percentage of these WDs always remains below 10\% and as shown in 
Hansen et al. (2007) the majority will not have had their stellar clocks affected sufficiently 
to affect the resulting WD luminosity function. 
One area where divorced WD contamination may be an issue is in the detection of 
post-cataclysmic variables which are expected to lie slightly off the standard WD 
sequence owing to residual heat from the mass-accretion process 
(e.g. Townsley \& Bildsten 2002). 
Looking at  Figures~\ref{f:fig3} and  \ref{f:fig4} we see that divorced WDs, 
which are similarly affected by mass accretion, could present confusion 
amongst such detections.

\section{Discussion}
\label{s:disc}

Considering that observed data is generally only obtained for a portion of 
a globular cluster, and that the location of the portion with respect to the 
cluster center may vary from study-to-study and cluster-to-cluster, it is 
important to understand how the local MF of a cluster varies with distance 
from the cluster center and how this compares to the global MF. 
We have shown that the local MF provides a good match to the global MF at, 
or just outside of, the half-mass radius of a cluster. 
This result is true at any age, including post-core-collapse. 
An earlier study by Vesperini \& Heggie (1997) also found that the MF in the 
intermediate region of a cluster provides a good resemblance to the global MF 
throughout the entire evolution. 
In this work the results were primarily based on $N$-body simulations with $N = 4\,096$ 
but with the claim that with careful scaling of the simulation time by the relaxation 
timescale the results could be taken to represent clusters starting with 
$6.15 \times 10^4 \, M_\odot$ in stars. 
This is comparable to the actual initial mass of our simulation and indicates that 
the Vesperini \& Heggie (1997) results do scale well to larger $N$ (or $M$). 
More recently Baumgardt \& Makino (2003) have looked at MF evolution based 
on $N$-body models with the initial $N$ ranging from $8\,192$ to $131\,072$ stars. 
They also found that the intermediate region, and in particular the region just 
outside of the half-light radius, gives the best agreement with the global MF. 

Interior to the half-mass radius the local MF becomes dominated by massive stars 
and has a flatter (or even reversed) slope compared to the global MF. 
Conversely, measuring the local MF in the halo of a cluster produces a steeper 
slope than that of the global MF. 
As discussed by Vesperini \& Heggie (1997) the evolution of the MF is a 
competition between mass-segregation and the escape (or evaporation) 
of stars owing to relaxation. 
The former effect flattens the MF in the inner regions of the cluster as it evolves 
while steepening the MF in the outer regions. 
However, the loss of stars from the outer region will act to flatten the MF so that 
the actual appearance of the MF in this region will depend on the relative 
efficiency of the two processes. 
Comparing the IMF and the local MF in the outermost shell of Figure~\ref{f:fig6} 
we see that the slopes are similar, with some flattening at the low-mass end. 
So at this age ($15\,$Gyr: near core-collapse) the escape of stars is only 
slightly more efficient than the process of mass-segregation. 
Looking at the cluster later (Figure~\ref{f:fig7}) the MF of the outermost shell 
is significantly flattened showing that the escape of stars is dominating as the 
cluster moves closer to dissolution. 
In Figure~\ref{f:fig8} we have attempted to give some indication of what is the 
best region of a cluster to observe in order to determine the IMF. 
In reality, owing to the dynamical processes just discussed, the present day MF 
quickly loses memory of the IMF (see also Vesperini \& Heggie 1997). 
Certainly by the time a cluster has reached core-collapse there is little hope 
of inferring the IMF unless the intervening action of the cluster dynamics is 
properly modeled. 
In the past it has often been the practice to fit multi-mass King-Michie models to the 
observed mass (or luminosity) function in order to correct for the effects of 
mass-segregation. 
However, these models assume energy equipartition and while clusters tend towards 
such a state it has been shown with $N$-body models that energy equipartition is 
only reached at late stages in the evolution, if at all in the outer regions 
(Giersz \& Heggie 1996; Baumgardt \& Makino 2003). 
We tend to agree with Baumgardt \& Makino (2003) that the effect of mass segregation, 
which is driven by energy equipartition, is best modeled by dynamical models. 

A distinction between our models and those presented previously by 
Vesperini \& Heggie (1997) and Baumgardt \& Makino (2003), for example, is 
that we have included primordial binaries. 
Looking at the radial variation of the MF in our 0\% binary reference model we 
do find that the local and global MFs generally match slightly nearer to the cluster 
center than they do for the 5\% binary models at the same age. 
This makes some sense owing to the difference in dynamical ages at the same 
physical age 
-- models with primordial binaries loses stars at a faster rate and thus have a shorter 
relaxation timescale compared to models without primordial binaries. 
As a result mass-segregation is more developed in the center of a primordial binary 
model and one must look further out to get a match to the global MF 
(or the IMF: the effect is noticeable in Figure~\ref{f:fig8}). 
However, observing near the half-mass radius is still optimal for recovering the 
global MF of an evolved star cluster, with or without binaries. 
The radial variation of the MF in our model with 10\% primordial binaries 
is indistinguishable from that of the 5\% model. 

Previous studies have demonstrated that the evolution of the mass function slope of 
a star cluster is strongly influenced by the position of a cluster with respect to the 
Galactic center (Vesperini \& Heggie 1997; Baumgardt \& Makino 2003). 
This is a result of the evaporation of stars across the tidal boundary owing to 
two-body relaxation proceeding at a faster rate for clusters at smaller $R_{\rm gc}$. 
We have touched on this topic in Section~\ref{s:method} when 
comparing the evolution timescales of model clusters at $R_{\rm gc}$ of 4 and $8.5\,$kpc. 
The issue will be investigated further in the near future when we present updated 
models for comparison to NGC$\,6397$ (see below). 
However, Baumgardt \& Makino (2003) have demonstrated that scaling model ages  
by their half-mass relaxation times removes the $R_{\rm gc}$ scaling of the 
MF evolution. 
It is also interesting to note that Vesperini \& Heggie (1997) found that the initial 
concentration of a cluster caused only a slight change in the two-body relaxation 
driven evaporation rate and thus had little effect on the MF slope. 
This is an important point considering that our current model has a lower core density than 
large core-collapse clusters such as NGC$\,6397$. 
Baumgardt \& Makino (2003) also showed that the density profile of a cluster was not 
a major factor in the MF evolution. 
Furthermore, they found that the MF behaviour could be scaled with $N$ provided that 
models were compared at the same fractional lifetime. 
So we can be fairly comfortable comparing our model to the NGC$\,6397$ data at 
similar dynamical ages. 

An effect that does have a potential bearing on the observed MF slope is disk-shocking. 
This has not been included in our model. 
Disc shocking has been shown by Vesperini \& Heggie (1997) to flatten the MF slope 
for clusters near the Galactic center ($R_{\rm gc} \le 4\,$kpc) but causes only minor 
differences for clusters on more distant orbits ($R_{\rm gc} \ge 8\,$kpc). 
As noted by Vesperini \& Heggie (1997) disk shocking is a mass-independent process 
in that the velocity change induced for a particular star depends only on the distance 
from the cluster center and not on the stellar mass. 
This means that disk shocking alone cannot alter the MF slope but in combination 
with mass segregation it can have a noticeable effect. 
The tidal shocks produced by frequent disk passages are certain to have played a 
role in the dynamical evolution of NGC$\,6397$ 
-- the disk crossing time is $100\,$Myr (Kalirai et al. 2007) which is comparable to 
the current half-mass relaxation time of $300\,$Myr (Harris 1996). 
This is an aspect we will be 
aware of when presenting future models for comparison with this cluster. 
Driven in part by the computational constraints that have kept $N$ small there 
has historically been little need for more than a simplistic treatment of 
tidal fields in $N$-body codes. 
Circular orbits about a point-mass galaxy and with an absence of disk-shocking 
are adequate for comparison to open clusters that reside in the Galactic disk. 
However, now that comparisons to globular cluster evolution are being considered, 
such a tidal field is a clear simplification. 
An obvious way to overcome this is to compute the orbit of the cluster in a 
realistic (and time-varying) Galactic potential and to do this in-step with modeling 
the cluster evolution. 
Such an approach will also be considered in future work. 

The results presented here and in Hurley et al. (2007), coupled with the binary frequency  
of 2\% measured by Davis et al. (2008b) for NGC$\,6397$,  indicate a very modest 
primordial binary content for the cluster. 
Indeed, Figure~\ref{f:fig5} shows that the binary fraction for MS stars may actually be 
an overestimate of the binary fraction for all stars in an evolved cluster. 
So 2\% may be an upper limit for the primordial binary frequency of NGC$\,6397$. 
However, there are other considerations. 
The distribution of orbital separations of the primordial binaries in the $N$-body model 
was capped at $100\,$au. 
If distributions of orbital parameters derived for field binaries, such as those of 
Duquennoy \& Mayor (1991) showing periods extending to $10^7\,$d, are taken 
as relevant to binaries formed in globular clusters (which is not necessarily the case) 
then this means that a number of wide binaries have been neglected from the starting model. 
These neglected binaries will be loosely bound, or ``soft'', and quickly broken-up in 
encounters with other stars or binaries (Heggie 1975). 
Therefore they will be of little consequence to the dynamical evolution of the cluster but 
they will impact the primordial binary fraction inferred from later values. 
The effect is discussed more fully in Hurley et al. (2007) and the worst-case scenario has 
the true primordial binary fraction being twice that used in the model. 
So the primordial binary frequency of NGC$\,6397$ could have been closer to 5\% but 
certainly nowhere near as high as the 100\% that has been proposed to explain 
current observations of the core binary fractions of clusters such as 47 Tucanae 
(Ivanova et al. 2005 -- although there are differences in the models presented in that study 
and here that need to be resolved). 
We note that our model starting with 10\% primordial binaries has a binary frequency of 9\% 
in the outer field at $15\,$Gyr. 

It is important to realize that none of the $N$-body models utilized in this work are being put forward 
as specific models of NGC$\,6397$. 
Instead we are using models that combine dynamical, stellar and binary evolution processes to aid our interpretation of the HST data for this cluster. 
NGC$\,6397$ is a core-collapsed cluster so we have chosen to focus on our simulated cluster at 
$15-16\,$Gyr, when it first reaches core-collapse, and up to $18\,$Gyr. 
Beyond this point the number of stars remaining decreases to levels where it becomes difficult  
to draw statistically significant inferences, especially when looking at radial variations within a model. 
We are not claiming that any of these times correspond to the age of NGC$\,6397$ and ideally we would have a post-core-collapse model at $3-4\,$Gyr earlier for comparison. 
However for old clusters such a difference is not substantial in stellar evolution terms 
(mass-loss rates and the main sequence turn-off mass, for example) so a comparison between 
the model and the data remains useful, especially when trying to understand the dynamical 
state of the cluster. 
To produce a model that exhibits core-collapse at an earlier age can be achieved 
by placing the cluster on an orbit closer to the Galactic center. 
As we have demonstrated this brings forward the onset of core-collapse but also 
leads to earlier cluster dissolution. 
Therefore, a corresponding increase in $N$ is also required in order to have 
stars remaining at the age of interest. 

Ideally we would perform a direct $N$-body model of NGC$\,6397$ with 
star-to star correspondence between the real and model cluster as has 
been done for open clusters such as M67 (Hurley et al. 2005). 
However, even when using special-purpose hardware such as the GRAPE 
boards the $N^3$ scaling of the $N$-body problem makes this computationally 
unfeasible. 
Consider that the $N = 100\,000$ simulations each took between $2$--$6\,$months 
to complete (with those orbiting at $4\,$kpc being the quickest) and had at most 
$1.6 \times 10^4 \, M_\odot$ of stars remaining at the age of NGC$\,6397$. 
This is a factor of 5--10 less than the estimated current mass of the cluster 
(Drukier 1995; Gnedin \& Ostriker 1997). 
To evolve a cluster on an orbit at $3$--$4\,$kpc from the Galactic center and have this 
much mass remaining at about $12\,$Gyr would require a starting model 
with $N \sim 400\,000$, as a rough estimate. 
This is clearly outside the scope of current $N$-body models and will only be realised 
with future hardware advances. 

Currently we have started an {\tt NBODY4} simulation with $200\,000$ stars and 2\% 
primordial binaries on an orbit at $4\,$kpc from the Galactic center 
(closer to the true orbit of NGC$\,6397$). 
This is projected to experience core-collapse at an age of $11$--$12\,$Gyr with 
$\sim 40\,000$ stars remaining but will also take up to a year to complete 
on a single 32-chip GRAPE-6 board. 
We note that while this improves the timescale problem it will still not be a direct model 
of NGC$\,6397$ and scaling issues related to the mismatch of $N$, such as 
differences in the central density and velocity dispersion, will remain. 
However, when completed, this model in combination with the NGC$\,6397$ HST data 
will further enhance our understanding of globular cluster evolution.

\section{Summary}
\label{s:summ}

We have presented $N$-body models of star cluster evolution created with 
the {\tt NBODY4} code. 
A range of models each starting with $100\,000$ stars and/or binaries have 
been considered with the proportion of primordial binaries (0, 5 or 10\%) and 
the galactocentric radius of the cluster orbit ($4$ or $8.5\,$kpc) 
being the parameters varied across the models. 
The model with 5\% primordial binaries on a circular orbit at $8.5\,$kpc from 
the Galactic center is the main focus of this work. 
This model reached core-collapse at an age of $16\,$Gyr and was terminated 
at $20\,$Gyr  when 9\% of the initial cluster mass remained bound. 

To mimic observations of the globular cluster NGC$\,6397$ we have contrasted 
the appearance of the model cluster in the center of the cluster with that in 
a region just outside of the half-mass radius. 
We look in detail at the evolution of the mass function and show that the local 
mass function of main-sequence stars near the half-mass radius is a 
good representation of the global mass function of a post-core-collapse cluster. 
The development of mass segregation is demonstrated 
-- our models complement previous work in this area that has been 
performed without the inclusion of primordial binaries. 
The mass functions and color-magnitude diagrams of the model are 
a good match to observations of NGC$\,6397$. 
We have also confirmed that the binary fraction of 0.02 observed near the half-mass 
radius of NGC$\,6397$ can be taken as representative of the primordial 
binary fraction of the cluster. 

We have discussed the validity of comparing a non-direct model with 
observations of NGC$\,6397$ in terms of the number of stars, the stellar 
evolution age, the density of stars and position within the Galaxy. 
Future models will be performed with twice the number of stars and will 
look at incorporating a non-idealized model of the Galactic potential as 
well as the effects of disk shocking.

\acknowledgments

JRH would like to thank the Swinburne RDS scheme for travel support during 
this work. JRH and MMS acknowledge the generous support of the Cordelia Corporation 
and that of Edward Norton which has enabled AMNH to purchase GRAPE-6 boards 
and supporting hardware. We also thank David Zurek and John Ouellette for 
maintaining these boards. 
HBR is grateful to the U.S.-Canada Fulbright Fellowship Committee, NSERC and 
the University of British Columbia (UBC) for support. 
DSD thanks the UBC-UGF for funding. 
IRK, JSK, BMSH, JA and RMR received support  from NASA/HST through grant GO-10424 
and JSK thanks NASA for support through a Hubble Fellowship. 
We thanks the referee for suggestions that aided this work.

\clearpage

\clearpage

\clearpage

\clearpage

\clearpage

\begin{deluxetable}{rrcrccrc}
\tablecolumns{8}
\tablewidth{0pc}
\tablecaption{
Summary of cluster parameters at $2\,$Gyr intervals for the main $N$-body model. 
\label{t:table1}
}
\tabletypesize\normalsize
\tablehead{ 
 $t$ & $N$ & $f_{\rm b}$ & $M_{\rm t}$ & $r_{\rm c}$ & $r_{\rm h}$ & $t_{\rm rh}$ & $\sigma$ 
}
\startdata
         0.0 & 100000 & 0.050 & 51910 & 2.72 & 6.66 & 1400.0 & 3.63 \\
  2000.0 &   84510 & 0.048 & 32830 & 3.40 & 8.58 & 2215.0 & 2.67 \\
  4000.0 &   74660 & 0.048 & 28455 & 3.06 & 8.56 & 2120.0 & 2.51 \\
  6000.0 &   64440 & 0.048 & 24845 & 2.63 & 8.15 & 1850.0 & 2.40 \\
  8000.0 &   54450 & 0.049 & 21590 & 2.21 & 7.50 & 1505.0 & 2.34 \\
10000.0 &   44910 & 0.051 & 18525 & 1.61 & 6.89 & 1200.0 & 2.26 \\
12000.0 &   36150 & 0.052 & 15670 & 0.97 & 6.10 &   905.0 & 2.20 \\
14000.0 &   28100 & 0.053 & 12925 & 0.68 & 5.44 &   670.0 & 2.13 \\
16000.0 &   20710 & 0.054 & 10140 & 0.17 & 4.97 &   490.0 & 1.96 \\
18000.0 &   14190 & 0.053 &  7405 & 0.34 & 4.77 &  390.0 & 1.76 \\
20000.0 &     8950 & 0.055 & 4895 & 0.17 & 4.12 &    250.0 & 1.55 \\
\enddata
\tablecomments{
Columns show the cluster age (Myr), number of objects (stars and binaries), 
binary fraction, total mass ($M_\odot$), core-radius (pc), half-mass radius (pc), 
half-mass relaxation timescale (Myr) and velocity dispersion (${\rm km} \, {\rm s}^{-1}$).  
}
\end{deluxetable}

\end{document}